\documentclass[a4paper,11pt]{article}
\pdfoutput=1 

\usepackage{jheppub} 

\usepackage[T1]{fontenc} 

\usepackage{subfig}

\newcommand{\m}{M_{H^{\pm}}}
\newcommand{\g}{\,\mbox{GeV}}
\newcommand{\la}{\lambda_1}
\newcommand{\lb}{\lambda_2}
\newcommand{\lc}{\lambda_3}

\newcommand{\lp}{\lambda_5}

\newcommand{\lczp}{\lambda_{345}}
\newcommand{\rg}{R_{\gamma\gamma}}

\newcommand{\fr}{\frac}
\newcommand{\relic}{\Omega_{DM}h^2}

\title{\boldmath Constraining Inert Dark Matter by $R_{\gamma\gamma}$ and WMAP data}

\author{Maria Krawczyk,}
\author[1]{Dorota Soko{\l}owska,\note{Corresponding author.}}
\author{Pawe{\l} Swaczyna}
\author{and Bogumi{\l}a \'{S}wie\.{z}ewska}

\affiliation{University of Warsaw, Faculty of Physics,\\Ho\.{z}a 69, 00-681 Warszawa, Poland}

\emailAdd{maria.krawczyk@fuw.edu.pl}
\emailAdd{dorota.sokolowska@fuw.edu.pl}
\emailAdd{pawel.swaczyna@student.uw.edu.pl}
\emailAdd{bogumila.swiezewska@fuw.edu.pl}

\abstract{We discuss the constraints on  Dark Matter coming from the LHC Higgs data and WMAP relic density measurements  for the Inert Doublet Model, which is  one of the simplest extensions of the Standard Model  providing a Dark Matter candidate.  We found that  combining   the  diphoton rate $\rg$ and the $\Omega_{DM}h^2$ data  one can set strong limits on the parameter space of the Inert Doublet Model, stronger or comparable to the constraints provided by the XENON100 experiment for low and medium Dark Matter mass.}
\arxivnumber{1305.6266}

\begin{document} 
\maketitle
\flushbottom

\section{Introduction}

Dark Matter (DM) is thought to constitute around 25\% of the Universe's mass-energy density, but its  precise nature is yet unknown. The DM relic density $\relic$ is well measured by WMAP and Planck experiments
and  the current value of $\relic$ is \cite{Beringer:2012}:
\begin{equation}
\Omega_{DM}h^2=0.1126 \pm 0.0036. \label{omega}
\end{equation}

Various direct and indirect 
detection experiments have reported signals that can be interpreted as DM particles. Low DM masses $\lesssim 10$ GeV are favoured by DAMA/LIBRA \cite{Bernabei:2008yi}, CoGeNT \cite{Aalseth:2010vx,Aalseth:2011wp} and recently by CDMS-II \cite{Agnese:2013rvf} experiment, while the medium mass region of 25 -- 60~GeV by CRESST-II \cite{Angloher:2011uu}. All those events lie in the regions excluded by the XENON10 and XENON100 experiments, which set the strongest limits on the DM-nucleon scattering cross-section \cite{Aprile:2012nq}. There have also been reports of the observation of products of the annihilation of DM particles, including the recent $130$ GeV  $\gamma$-line from the Fermi-LAT experiment \cite{Bringmann:2012vr, Weniger:2012tx, Fermi-LAT:2013uma}. However, there is no agreement as to whether one can truly interpret those indirect measurements as a proof of existence of Dark Matter (see e.g. \cite{Bringmann:2012ez, Bergstrom:2012fi} for reviews).

There have been many attempts to explain those contradictory results  either by assuming some experimental inaccuracies coming from incorrectly determined physical quantities in astrophysics or nuclear physics, or by interpreting the results in modified  astrophysical  models of DM  (see e.g. \cite{Collar:2010gg,XENON100:2010dsa,Collar:2010gd, Kopp:2011yr,Frandsen:2013cna}). However, so far no agreement has been reached, and the situation in direct and indirect detection experiments is not yet clear \cite{Kopp:2011yr, Bergstrom:2012fi,Savage:2008er, Cline:2012ei}.

In this paper we set constraints on the scalar DM particle from the Inert Doublet Model (IDM), using solely  the LHC Higgs data and relic density measurements. The IDM provides an example of a Higgs portal DM. In a vast region of the allowed DM masses, particularly in the range that the LHC can directly test, the main annihilation channel of DM particles and their interaction with nucleons, relevant for direct DM detection, are processed by exchange of the Higgs particle. We found that the $h \to \gamma \gamma$ data for the SM-like Higgs particle with mass $M_h \approx(125-126)$ GeV  sets strong constraints on the allowed masses and couplings of DM in the IDM. Combining them with the WMAP results excludes a large part of the IDM parameter space, setting limits on DM that are stronger or comparable to those obtained by XENON100.
 
\section{Inert Doublet Model}

The Inert Doublet Model is defined as a 2HDM with an exact $D$ ($Z_2$ type) symmetry: $\phi_S \to \phi_S, \phi_D \to - \phi_D$ \cite{Cao:2007rm,Barbieri:2006dq}, i.e. a 2HDM with a $D$-symmetric potential, vacuum state and Yukawa interaction (Model I). In the IDM  only one doublet, $\phi_S$, is involved in the Spontaneous Symmetry Breaking, while the $D$-odd doublet, $\phi_D$, is inert, having $\langle \phi_D \rangle =0$ and no couplings to fermions. The lightest particle coming from this doublet is stable, being a good Dark Matter candidate. 

The IDM provides, apart from the DM candidate, also a good framework for  studies of the thermal evolution of the Universe \cite{Krawczyk:2010,Sokolowska:2011aa,Sokolowska:2011sb,Sokolowska:2011yi},  electroweak symmetry breaking \cite{refTytEW},  strong electroweak phase transition \cite{Kanemura:2004ch,Gil:2012-ewpt,Cline:2013,Chowdhury:2011ga} and neutrino masses \cite{Ma:2006km,Gustafsson:2012vj}.

The $D$-symmetric  potential of the IDM has the following form:
\begin{equation}\begin{array}{c}
V=-\fr{1}{2}\left[m_{11}^2(\phi_S^\dagger\phi_S)\!+\! m_{22}^2(\phi_D^\dagger\phi_D)\right]+
\fr{\lambda_1}{2}(\phi_S^\dagger\phi_S)^2\! 
+\!\fr{\lambda_2}{2}(\phi_D^\dagger\phi_D)^2\\[2mm]+\!\lambda_3(\phi_S^\dagger\phi_S)(\phi_D^\dagger\phi_D)\!
\!+\!\lambda_4(\phi_S^\dagger\phi_D)(\phi_D^\dagger\phi_S) +\fr{\lambda_5}{2}\left[(\phi_S^\dagger\phi_D)^2\!
+\!(\phi_D^\dagger\phi_S)^2\right],
\end{array}\label{pot}\end{equation}
with all  parameters real
(see e.g. \cite{Krawczyk:2010}). 
The vacuum state in the IDM is given by:\footnote{In a 2HDM with the potential $V$ (\ref{pot}) different vacua can exist, e.g. a mixed one with $\langle\phi_S\rangle\neq0$, $\langle\phi_D\rangle\neq0$ or an inertlike vacuum with $\langle\phi_S\rangle=0$, $\langle\phi_D\rangle\neq0$, see~\cite{Krawczyk:2010,Sokolowska:2011aa,Sokolowska:2011sb,Sokolowska:2011yi}.}
\begin{equation}
\langle\phi_S\rangle =\frac{1}{\sqrt{2}} \begin{pmatrix}0\\ v\end{pmatrix}\,,\qquad \langle\phi_D\rangle = \frac{1}{\sqrt{2}}
\begin{pmatrix} 0 \\ 0  \end{pmatrix}, \quad v = 246 \textrm{ GeV}.\label{dekomp_pol}
\end{equation}
The first doublet, $\phi_S$, contains the  SM-like Higgs boson $h$ with mass $M_h$ equal to 
\begin{equation}
M_{h}^2=\lambda_1v^2= m_{11}^2 = \left(125 \textrm{ GeV}\right)^2. \label{Higgsmass}
\end{equation}
The second doublet, $\phi_D$, consists of four dark (inert) scalars $H,\,A,\,H^\pm$, which do not 
couple to fermions at the tree-level.
Due to an exact $D$ symmetry the lightest neutral scalar $H$ (or $A$) is stable and can play a role of the DM.\footnote{Charged DM in the IDM is excluded by the interplay between perturbativity and positivity constraints~\cite{Krawczyk:2010}.} The masses of the dark particles read:
\begin{equation}\begin{array}{c}
M_{H^\pm}^2=\fr{1}{2} \left(\lambda_3 v^2-m_{22}^2\right)\,,\\[3mm]
M_{A}^2=M_{H^\pm}^2+\fr{1}{2}\left(\lambda_4-\lambda_5\right)v^2\,,\quad M_{H}^2=
M_{H^\pm}^2+\fr{1}{2}\left(\lambda_4+\lambda_5\right)v^2\,.
\end{array}\label{mass}\end{equation}
We take $H$ to be the DM candidate and so $M_H < M_A, M_{H^\pm}$ ($\lp<0,\ \lambda_{4}+\lambda_{5}<0$).

The properties of the IDM can be described by the parameters of the potential $m_{ii}^2$ and $\lambda_i$ or by the masses of the scalar particles and their physical couplings. The parameter $\lambda_{345}=\lambda_3 + \lambda_4 + \lambda_5$ is related to a triple and a quartic coupling between the SM-like Higgs $h$ and the DM candidate $H$, while $\lambda_3$ describes the Higgs particle interaction with charged scalars $H^\pm$. The parameter $\lambda_2$ gives the quartic self-couplings of dark particles.
Physical parameters are limited by various  theoretical and experimental constraints (see e.g.~\cite{Cao:2007rm, Agrawal:2008xz, Gustafsson:2007pc, Dolle:2009fn, Dolle:2009ft, LopezHonorez:2006gr, Arina:2009um, Tytgat:2007cv, Honorez:2010re, Krawczyk:2009fb, Kanemura:1993, Akeroyd:2000, Swiezewska:2012, Gustafsson:2009, Gustafsson:2010,Gustafsson:2012aj}). We take the following conditions into account:
\paragraph{Vacuum stability} We require that the potential is bounded from below, which leads to the following constraints~\cite{Nie:1998yn}:
\begin{equation}
\lambda_1>0,\quad\lambda_2>0,\quad\lc+\sqrt{\la\lb}>0,\quad\lczp+\sqrt{\la\lb}>0\,\,\,\, (\lczp=\lambda_3+\lambda_4+\lambda_5). \label{stability}
\end{equation}

These are tree-level positivity conditions, which ensure the existence of a global minimum. It is known that in the Standard Model the radiative corrections, mainly the top quark contribution, lead to negative values of the Higgs self-coupling, and thus to the instability of the SM vacuum for larger energy scales. The SM vacuum can be metastable, if its lifetime is long enough, i.e. longer than the lifetime of the Universe, see e.g. \cite{EliasMiro:2011aa}. An analysis of the stability of the potential in the IDM beyond the tree-level approximation is more complicated and it is beyond the scope of this paper. However, it has been shown in Ref.~\cite{Goudelis:2013uca} that in the IDM the contributions from four additional scalar states will in general lead to the relaxation of the stability bound, as compared to the SM. This allows the IDM to be valid (i.e. having a stable, and not a metastable vacuum) up to the Planck scale, for a wide portion of the parameter space of the IDM for the currently measured values of the Higgs boson and top quark masses.

\paragraph{Existence of inert vacuum} In the IDM two minima of different symmetry properties can coexist
\cite{Krawczyk:2010,Sokolowska:2011aa,Sokolowska:2011sb,Sokolowska:2011yi}. For the state (\ref{dekomp_pol}) to be not just a \textit{local}, but  the \textit{global} minimum,  the following  condition has to be fulfilled~\cite{Krawczyk:2010}:\footnote{In principle the IDM allows for tree-level metastability, if the inert minimum is a local one with a~lifetime larger than the age of the Universe. In such a case the inertlike minimum would be a true vacuum. However, for the sake of clarity in this work we limit ourselves only to a case in which inert is a global minimum.}
\begin{equation}
m_{11}^2/\sqrt{\la}>m_{22}^2/\sqrt{\lb}.\label{inertvac}
\end{equation}
\paragraph{Perturbative unitarity} Parameters of the potential are constrained by the following bound on the eigenvalues of the high-energy scattering matrix of the scalar sector: $|\Lambda_i|<8\pi$~\cite{Kanemura:1993,Akeroyd:2000,Swiezewska:2012}, which leads to the upper limit on the DM quartic self-coupling:
\begin{equation}
\lambda_{2}^{\textrm{max}} = 8.38. \label{unitarity}
\end{equation}
The value of the Higgs boson mass (\ref{Higgsmass}) and conditions (\ref{stability},\ref{inertvac},\ref{unitarity}) provide the following constraints 
\cite{Swiezewska:2012}:
\begin{equation}
 \lambda_1 = 0.258, \quad m_{22}^2\lesssim 9\cdot10^4\g^2, \quad  \lambda_3, \lambda_{345} > -\sqrt{\lambda_1\lambda_2} \geqslant -1.47. \label{constraints}
\end{equation}

\paragraph{EWPT} Values of the $S$ and $T$ parameters should lie within $2\sigma$ ellipses of the $(S,T)$ plane with the following central values~\cite{Nakamura:2010}: $S=0.03\pm0.09$, $T=0.07\pm0.08$, with correlation equal to 87\%. 

\paragraph{LEP limits} 
The LEP II analysis excludes the region of masses in the IDM where simultaneously~\cite{Gustafsson:2009,Gustafsson:2010}: 
\begin{equation}  \label{LEP}
M_{H} < 80  \textrm{ GeV},\ M_{A} < 100 \textrm{ GeV and } \delta_A = M_A-M_H > 8 \textrm{ GeV}.
\end{equation}
For $\delta_A<8$ GeV the LEP I limit applies \cite{Gustafsson:2009,Gustafsson:2010}:
\begin{equation}
M_{H} + M_{A} > M_Z. \label{LEPI}
\end{equation}
The standard limits 
for the charged scalar in 2HDM do not apply, as $H^\pm$ has no couplings to fermions. Its mass is indirectly constrained by the  studies of supersymmetric models at LEP to be \cite{Pierce:2007ut}:
\begin{equation}
\m \gtrsim 70-90 \textrm{ GeV}.
\end{equation}

\paragraph{Relic density constraints} In 
a big part of the parameter space of the IDM the value of $\Omega_{DM}h^2$ predicted by the IDM  is too low, meaning that $H$ does not constitute 100\% of DM in the Universe. However, there are three regions of $M_{H}$ in agreement with $\Omega_{DM}h^2$  (\ref{omega}): (i) light DM particles with mass $\lesssim 10 \textrm{ GeV}$, (ii) medium DM mass of $40-150 \textrm{ GeV}$ and (iii) heavy DM  with mass $\gtrsim 500 \textrm{ GeV}$. Proper relic density (\ref{omega}) can be obtained by tuning the $\lambda_{345}$ coupling, and in some cases also by the coannihilation between $H$ and other dark scalars and interference processes with virtual EW gauge bosons \cite{Cao:2007rm, Barbieri:2006dq, Gustafsson:2007pc, Dolle:2009fn, Dolle:2009ft, LopezHonorez:2006gr, Arina:2009um, Tytgat:2007cv, Honorez:2010re,LopezHonorez:2010tb,Sokolowska:2011aa,Sokolowska:2011sb}.
 

\section{$\rg$ constraints  for the dark scalars}


A SM-like Higgs particle was discovered at the LHC in 2012. 
$\rg$, the ratio of the diphoton decay rate of the observed $h$ to the SM prediction, is sensitive to the "new physics". 
The current measured values of $\rg$ provided by the ATLAS and the CMS collaborations are respectively \cite{ATLAS:2013oma,CMStalk}:
\begin{eqnarray}
\textrm{ATLAS} &: & \rg = 1.65\pm0.24\mathrm{(stat)}^{+0.25}_{-0.18}\mathrm{(syst)}, \label{rg_atlas} \\
\textrm{CMS} & : &  \rg = 0.79^{+0.28}_{-0.26}. \label{rg_cms}
\end{eqnarray}
Both of them are in 2$\sigma$ agreement with the SM value $\rg = 1$, however a deviation from that value is still possible and would be an indication of physics beyond the SM.
 
The ratio $\rg$ in the IDM is given by:
\begin{equation}\label{rgg}
R_{\gamma \gamma}:=\frac{\sigma(pp\to h\to \gamma\gamma)^{\textrm{IDM}}}{\sigma(pp\to h\to \gamma\gamma)^{\textrm  {SM}}}
\approx \frac{\Gamma(h\to \gamma\gamma)^{\mathrm{IDM}}}{\Gamma(h\to \gamma\gamma)^{\mathrm{SM}}}\frac{\Gamma(h)^{\mathrm{SM}}}{\Gamma(h)^{\mathrm{IDM}}} \, ,
\end{equation}
where $\Gamma(h)^{\mathrm{SM}}$ and $\Gamma(h)^{\mathrm{IDM}}$ are the total decay widths of the Higgs boson in the SM and the IDM respectively, while $\Gamma(h\to \gamma\gamma)^{\mathrm{SM}}$ and $\Gamma(h\to \gamma\gamma)^{\mathrm{IDM}}$ are the respective partial decay widths for the process $h\to\gamma\gamma$. In (\ref{rgg}) the facts that the main production channel is gluon fusion and that the Higgs particle from the IDM is SM-like, so $\sigma(gg\to h)^{\textrm{IDM}} = \sigma(gg\to h)^{\textrm{SM}}$, were used. In the IDM two sources of deviation from $\rg=1$ are possible.
First is a charged scalar contribution to the partial decay width $\Gamma(h\to \gamma\gamma)^{\textrm{IDM}}$  ~\cite{Cao:2007rm, Djouadi:2005, Djouadi:2005sm, Posch:2010, Arhrib:2012}:
\begin{equation}
 \Gamma(h \rightarrow \gamma\gamma)^{\mathrm{IDM}}=\frac{G_F\alpha^2M_h^3}{128\sqrt{2}\pi^3}\bigg|\underbrace{\frac{4}{3}A_{1/2}\left(\frac{4M_t^2}{M_h^2} \right)+A_1\left(\frac{4M_W^2}{M_h^2} \right)}_{\mathcal{M}^{\mathrm{SM}}}+\underbrace{\frac{\lambda_3 v^2}{2M_{H^\pm}^2}A_0 \left(\frac{4M_{H^\pm}^2}{M_h^2} \right)}_{\delta\mathcal{M}^{\mathrm{IDM}}}\bigg|^2\,, \label{Hloop}
\end{equation}
where $\mathcal{M}^{\textrm{SM}}$ is the SM amplitude and $\delta\mathcal{M}^{\textrm{IDM}}$ is the $H^\pm$ contribution.\footnote{The definition of the functions  $A_i$ can be found in refs.~\cite{Djouadi:2005, Djouadi:2005sm}.}
The interference between $\mathcal{M}^{\textrm{SM}}$ and $\delta\mathcal{M}^{\textrm{IDM}}$ can be either constructive or destructive, leading to an increase or a~decrease of the decay rate~(\ref{Hloop}).

The second source of  modifications of $\rg$  are the possible invisible decays $h\to HH$ and $h\to AA$, which can strongly augment the total decay width $\Gamma^{\textrm{IDM}}(h)$  with respect to the SM case. Partial widths for these decays  are given by:
\begin{equation}\label{inv-width}
 \Gamma(h\to HH)=\frac{\lambda_{345}^2v^2}{32\pi M_h}\sqrt{1-\frac{4M_{H}^2}{M_h^2}}\, , 
\end{equation}
with $M_H$ exchanged to $M_A$ and $\lambda_{345}$ to $\lambda_{345}^-$ ($\lczp^-=\lambda_3+\lambda_4-\lambda_5$), for the $h\to AA$ decay. Using eq.~(\ref{mass}) one can reexpress the couplings $\lc$ and $\lczp^-$ in terms of $M_H,\ M_A,\ \m$ and $\lczp$, and so from eq.~(\ref{Hloop}) and~(\ref{inv-width})
$\rg$   depends only on the masses of the dark scalars and $\lambda_{345}$.


For $M_H>M_h/2$ (and $M_A>M_h/2$)  the invisible channels are closed, and $\rg >1$ is possible, with the maximal value of $\rg$ equal to $3.69$ for $M_H = \m = 70$ GeV. 

\begin{figure}[t]
  \centering
    \includegraphics[width=.6\textwidth]{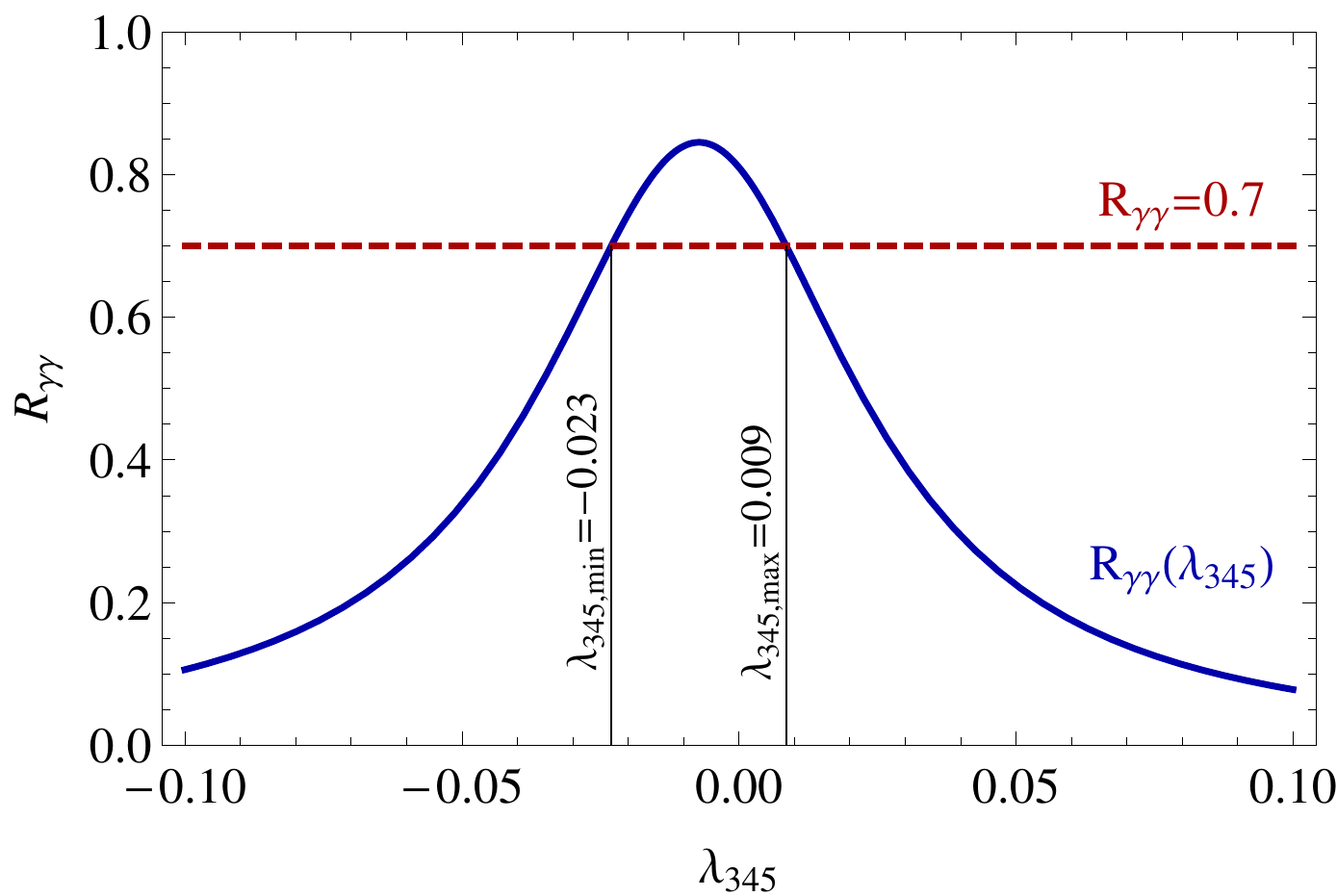} 
  \caption{ $\rg$ as a function of $\lczp$ for the following masses of dark scalars: $M_H=55\g,\ M_A=60\g,\ \m=120\g$. The bounds on $\lczp$ coming from the requirement that $\rg>0.7$ are shown.
  \label{rgg-l345}}
\end{figure}

If $M_{H} <M_h/2$ then the $h\to HH$ invisible channel is open and it is not possible to obtain $\rg>1$, as shown in \cite{Swiezewska:2012eh,Arhrib:2012}.  If an enhancement (\ref{rg_atlas}) in the diphoton channel is confirmed, this DM mass region is already excluded. However, if the final value of $\rg$ is below 1, as suggested by the CMS data (\ref{rg_cms}), then it limits the parameters of the IDM on the basis of the following reasoning.  For any given values of the dark scalars' masses $\rg$ is a function of one parameter: $\lczp$, the behaviour of which is presented in figure~\ref{rgg-l345} for $M_H=55\g,\ M_A=60\g,\ \m=120\g$ (the same shape of the curve is preserved for different values of masses). It can be observed, that setting a lower bound on $\rg$ leads to upper and lower bounds on $\lczp$. We will explore these bounds, as functions of $M_H$ and $\delta_A$ in Sections~\ref{sec-open} and~\ref{sec_Aclosed} for three cases that are in 1$\sigma$ region of the CMS value: $\rg>0.7,\,0.8,\,0.9$, 
respectively.





\subsection{$HH, AA$ decay channels open\label{sec-open}}

If both $M_H, M_A<M_h/2$ then the LEP constraint (\ref{LEP}) enforces $\delta_A<8\g$ and so eq.~(\ref{LEPI}) limits the allowed values of the DM particle mass 
$M_H>(M_Z - 8 \g)/2 \approx 41\g$.
 In this region, the invisible decay channels have stronger influence on the value of $\rg$  than the contribution from the charged scalar loop \cite{Swiezewska:2012eh}, and so the exact value of $\m$ influences the results less than the other scalar masses. In the following examples we use $\m = 120$~GeV, which is a good benchmark value of the charged scalar mass in the DM analysis for the low and medium DM mass regions, discussed later in section \ref{consequences}. Due to the  dependence of the partial width $\Gamma(h \rightarrow AA)$ on $|\lambda_{345}^-|$ the obtained lower and upper bounds are not symmetric with respect to $\lambda_{345}=0$.

\begin{figure}[t]
  \centering
  \subfloat[$R_{\gamma\gamma}>0.7$]{
    \includegraphics[width=.33\textwidth]{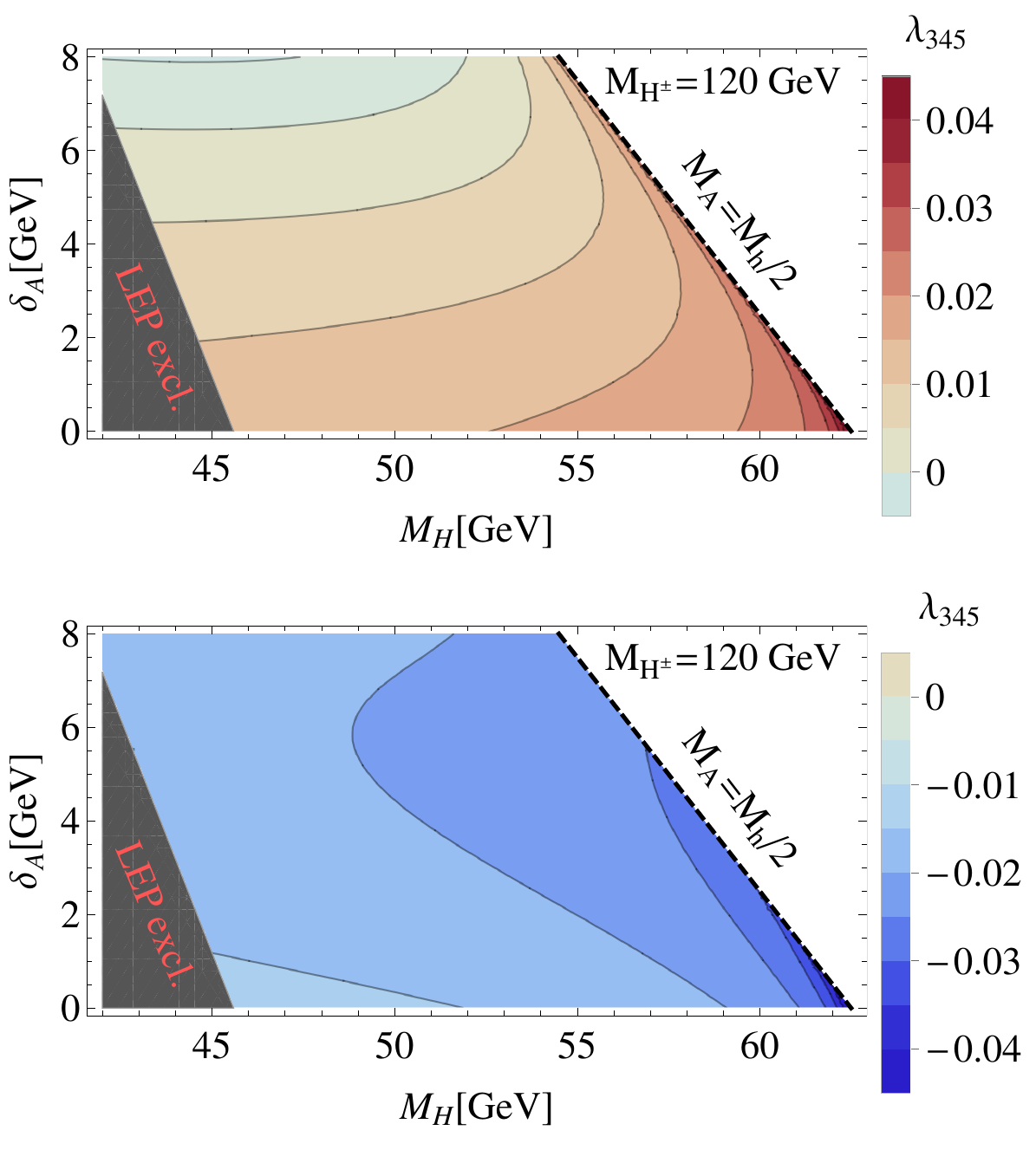} \label{r066}}
  \subfloat[$R_{\gamma\gamma}>0.8$]{
    \includegraphics[width=.33\textwidth]{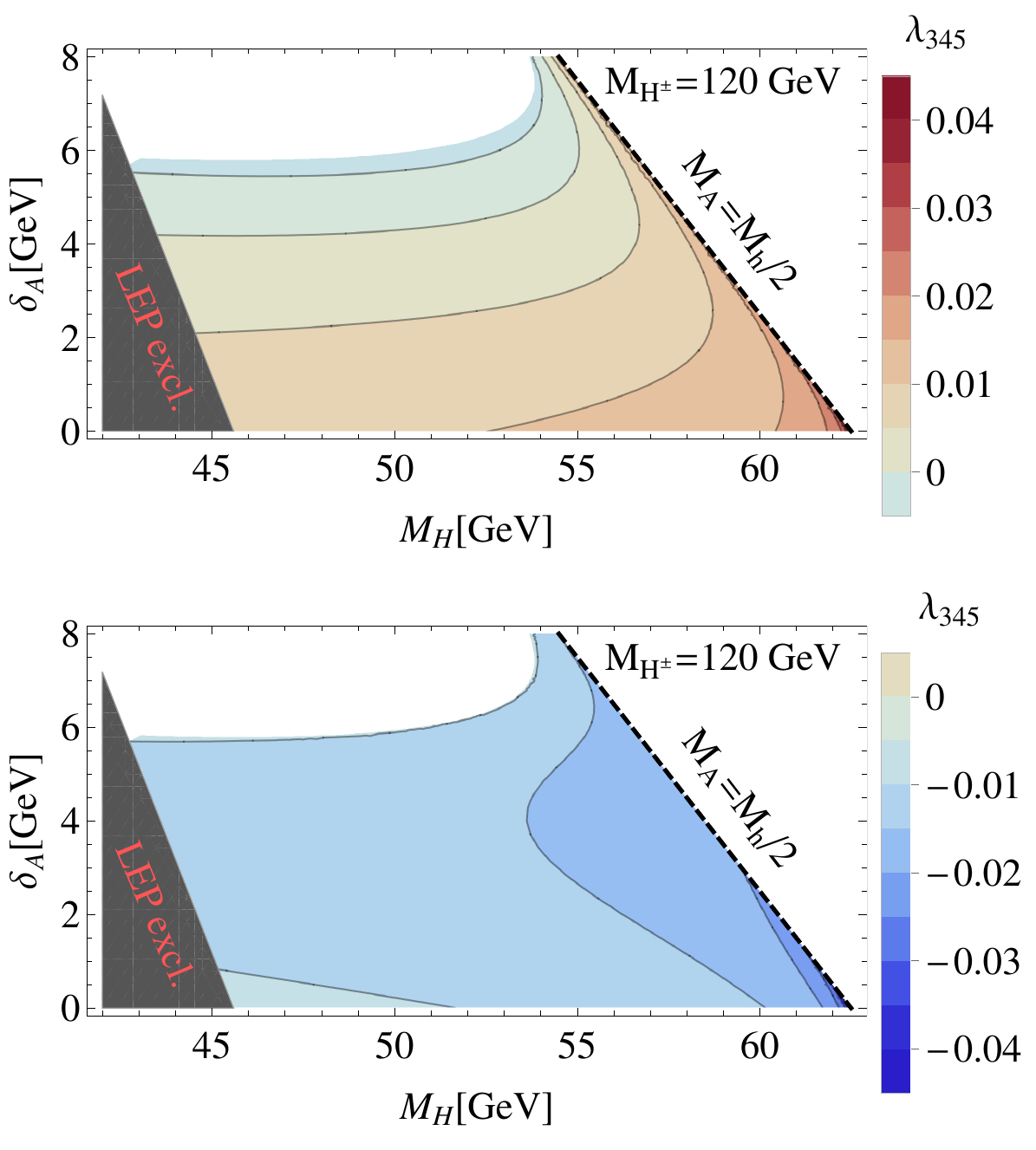} \label{r08}}
  \subfloat[$R_{\gamma\gamma}>0.9$]{
    \includegraphics[width=.33\textwidth]{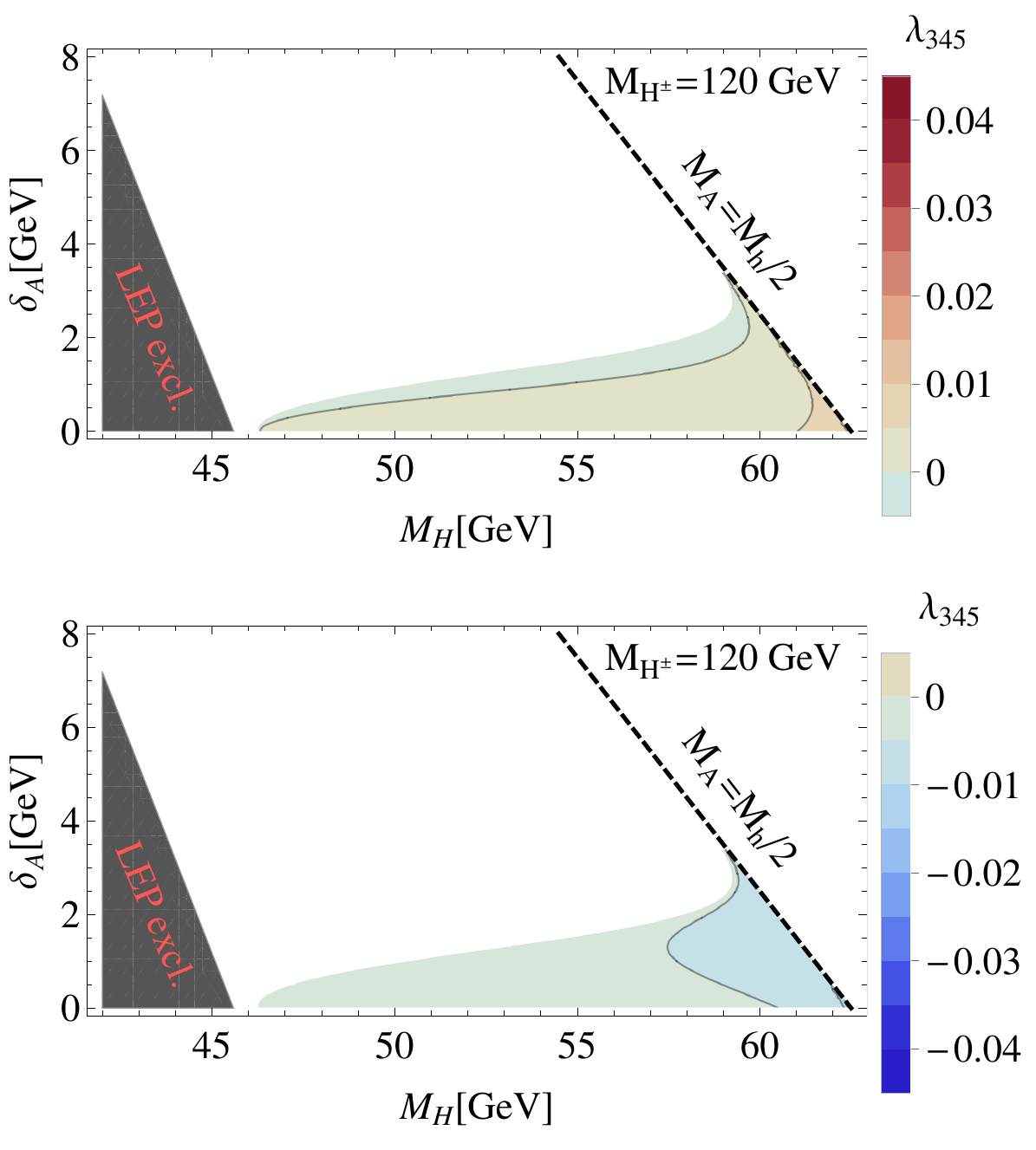} \label{r09}}
  \caption{Upper (upper panel) and lower (lower panel) limits on $\lambda_{345}$ coming from the requirement that (a) $\rg>0.7$, (b) $\rg>0.8$, (c) $\rg>0.9$, expressed as functions of $M_H$ and $\delta_A$ for the case when the $h\to HH,\, AA$ channels are open ($M_H,M_A<M_h/2$). $M_{H^{\pm}}$ is set to $120\g$. The lower left corner is excluded by LEP.}
  \label{rgg:fig:lam345open}
\end{figure}

\paragraph{Diphoton rate constraints} Figure \ref{r066} shows the upper and lower limits for the $\lambda_{345}$ coupling if $\rg > 0.7$. 
The allowed values of $\lambda_{345}$ are small, typically between $(-0.04,0.04)$, depending on the difference between masses of $H$ and $A$. 
In general, for $\rg > 0.8$ the allowed values of $\lambda_{345}$ are smaller than for $\rg >0.7$. Also, region of larger $\delta_A$ is excluded (figure \ref{r08}). 
 In contrast to the previous cases condition $\rg > 0.9$ strongly limits the allowed parameter space of the IDM, as shown in figure \ref{r09}, where a large portion of the  parameter space is excluded. The allowed  $A,H$ mass difference is $\delta_A \lesssim 2$ GeV, and values of $\lambda_{345}$ are smaller than in the previous cases. Requesting  larger $\rg$ leads to the exclusion of the whole region of masses, apart from $M_H \approx M_A \approx M_h/2$.

\begin{figure}[t]
  \centering
  \subfloat[$\textrm{Br}(h\to \textrm{inv}) < 65 \%$]{
    \includegraphics[width=.4\textwidth]{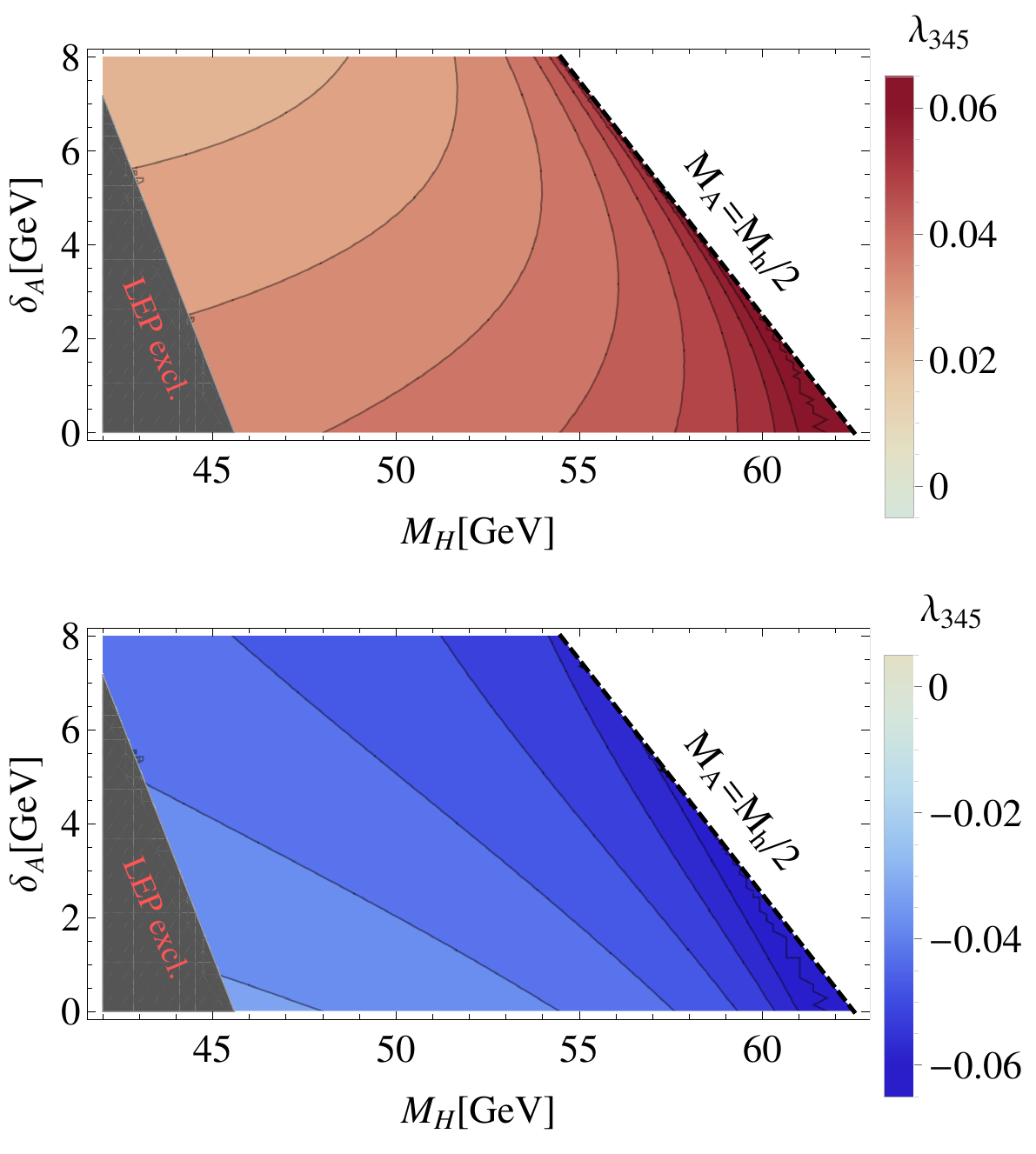} \label{inv65}}
  \subfloat[$\textrm{Br}(h\to \textrm{inv}) < 20 \%$]{
    \includegraphics[width=.4\textwidth]{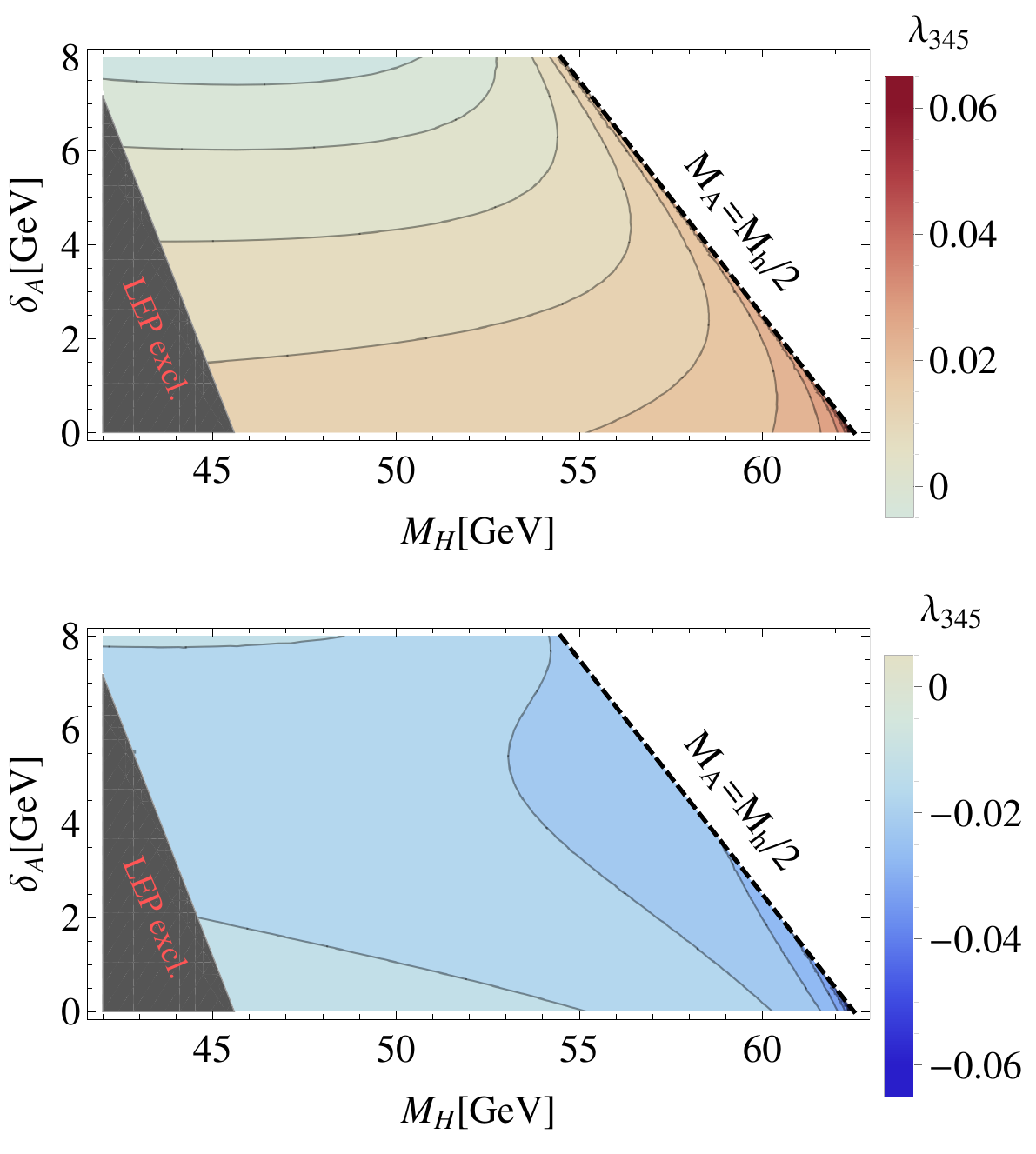} \label{inv20}}
    \caption{Upper (upper panel) and lower (lower panel) limits on
$\lambda_{345}$ coming from the requirement that (a)
$\mathrm{Br}(h\to\mathrm{inv})<65\%$, (b)
$\mathrm{Br}(h\to\mathrm{inv})<20\%$
expressed as functions of $M_H$ and $\delta_A$ for the case when the
$h\to HH,\, AA$ channels are open. The lower left corner is excluded by LEP.
} \label{inv}
\end{figure}

\paragraph{$\pmb{\textrm{Br}(h\to \textrm{inv})}$} In principle, while discussing the $M_H<M_h/2$ region, one should also include the constraints  from existing LHC data on the invisible channels branching ratio \cite{Djouadi:2012zc,Goudelis:2013uca}. However, constraints on $\lambda_{345}$ obtained by requesting $\textrm{Br}(h\to \textrm{inv}) < 65\, \%$ \cite{ATLAS:2013oma} are up to 50\% weaker than those coming from $\rg$, compare figure \ref{rgg:fig:lam345open}  and figure \ref{inv65}. The limits from the invisible branching ratio start to be comparative with the $\rg$ constraints when $\textrm{Br}(h\to \textrm{inv}) < 20\, \%$, as estimated in \cite{Belanger:2013kya,Dumont:2013mba}.



\subsection{$AA$ decay channel closed \label{sec_Aclosed}}

\begin{figure}[t]
\centering
\subfloat[$\rg>0.7$]{
\includegraphics[width=.33\textwidth]{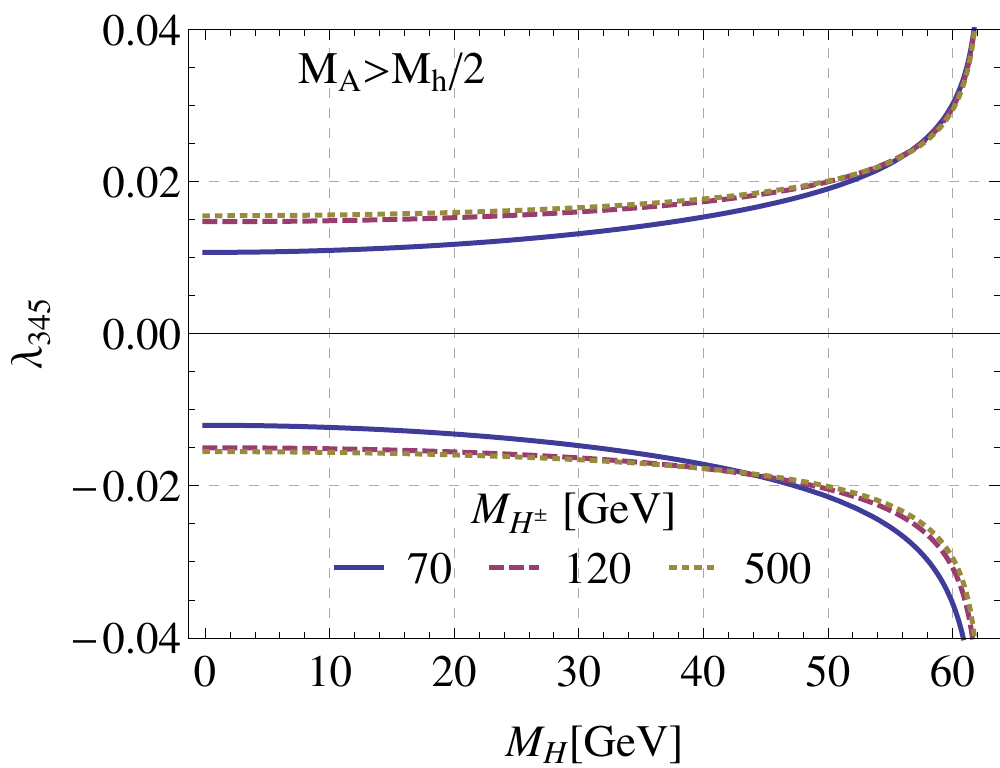}}
\subfloat[$\rg>0.8$]{
\includegraphics[width=.33\textwidth]{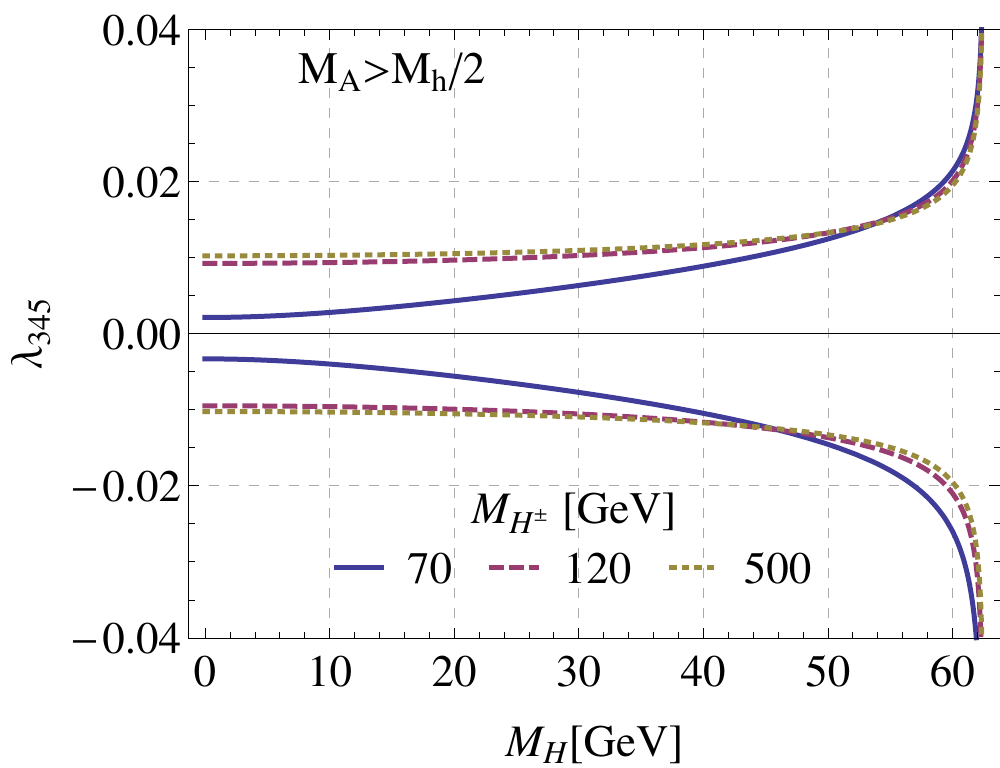}}
\subfloat[$\rg>0.9$]{
\includegraphics[width=.33\textwidth]{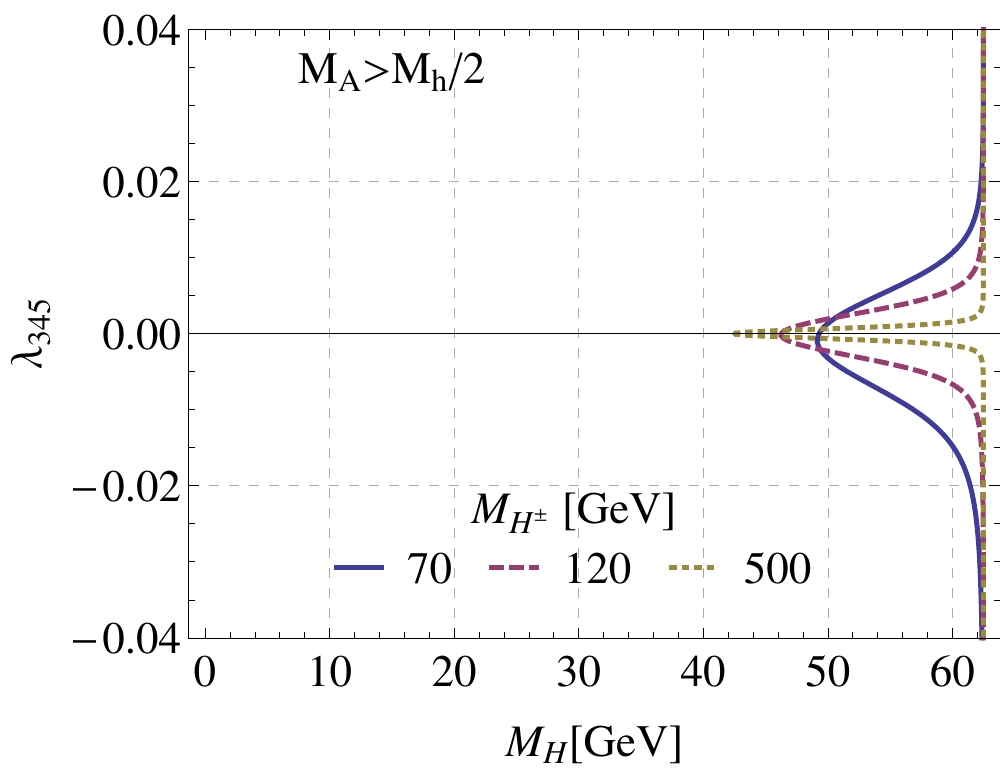}}

\caption{Upper and lower limits on $\lambda_{345}$ coming from the
requirement that (a) $\rg>0.7$, (b) $\rg>0.8$, (c) $\rg>0.9$, expressed
as functions of $M_H$, for the case when the $h\to AA$ channel is closed.
Three values of $M_{H^{\pm}}$ are considered $\m=70\g$, $\m=120\g$,
$\m=500\g$.} \label{Aclosed}
\end{figure}

When the $AA$ decay channel is closed, a very light DM particle can exist. Of course, if the $AA$ channel is closed the values of $\rg$ do not depend on the value of $M_A$,
while the charged scalar contribution becomes more relevant. A clear dependence on the $H^\pm$ mass appears especially for $\m \lesssim 120$ GeV. Figure \ref{Aclosed} shows the limits on $\lambda_{345}$ coupling that allow  values of $\rg$ higher than 0.7, 0.8 and 0.9 for $\m = 70,\,120$ and 500 GeV,  respectively.
Larger value of $\rg$ leads to  smaller allowed values of $\lambda_{345}$. In the case of $\rg>0.9$ a~large region of DM masses is excluded, as it is not possible to obtain the requested value of $\rg$ for any value of $\lambda_{345}$. 

If $\rg>0.7$ then an exact value of $\m$ is not crucial for the obtained limits on $\lambda_{345}$, and allowed values of $|\lambda_{345}|$ are of the order of $ 0.02$. For $\rg >0.8$ the obtained bounds are clearly different for $\m = 70$ GeV and 120 GeV. Smaller $H^\pm$ mass leads to stronger limits, requiring $|\lambda_{345}|\sim 0.005$, while larger masses of $H^\pm$ allow $|\lambda_{345}| \sim 0.015$. 

Condition $\rg >0.9$ limits the IDM parameter space  strongly. It is not possible to have $\rg >0.9$ if $M_H \lesssim 45$ GeV. For the larger masses only relatively small values of $\lambda_{345}$ (below 0.02) are allowed. It is interesting to note, that in this case not smaller, but larger $\m$ leads to more stringent limits on $\lambda_{345}$.

\begin{figure}[t]
\centering
\includegraphics[width=.6\textwidth]{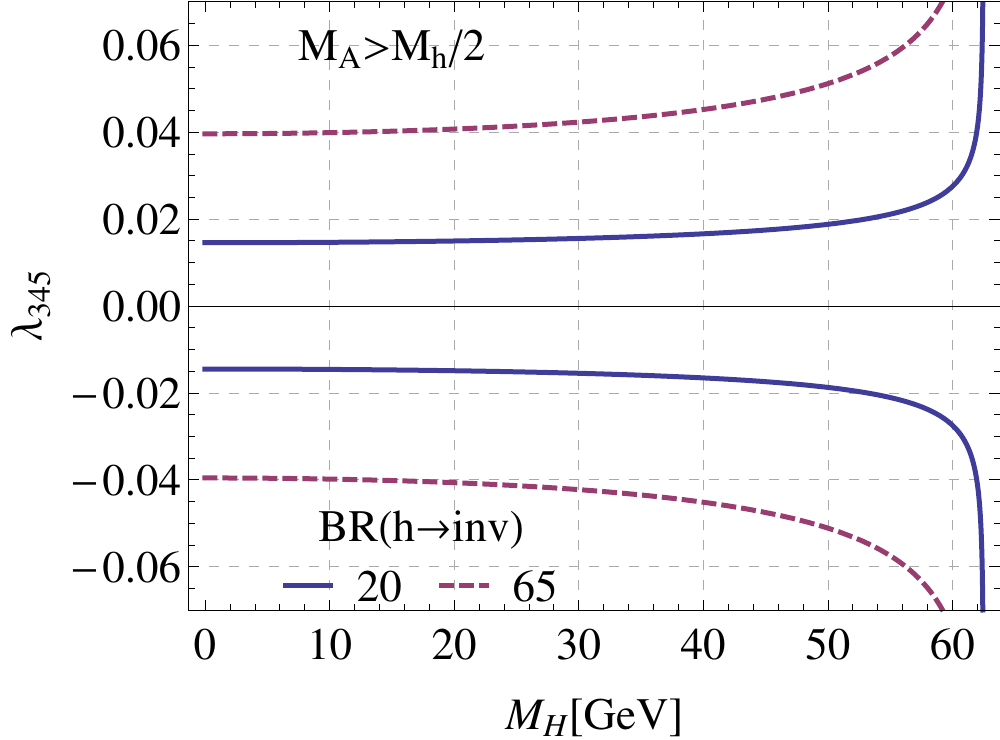}
\caption{Upper and lower limits on $\lambda_{345}$ coming from the
bounds on the branching ratio: $\mathrm{Br}(h\to\mathrm{inv})<65\%$ (dashed line) and $\mathrm{Br}(h\to\mathrm{inv})<20\%$ (solid line), expressed as a
function of $M_H$, for the case when the $h\to AA$ channel is closed.} \label{invclosed}
\end{figure}

\paragraph{$\pmb{\textrm{Br}(h\to \textrm{inv})}$} Similarly to the $M_A<M_h/2$ case, the constraint from the invisible decay branching ratio $\mathrm{Br}(h\to\mathrm{inv})<65\%$ does not further limit the values of $\lambda_{345}$, 
compare figure \ref{Aclosed} and figure \ref{invclosed}. Bounds obtained from $\mathrm{Br}(h\to\mathrm{inv})<20\%$ are competitive with those coming from $\rg$.

\paragraph{DM-nucleon cross section}

In the IDM the DM-nucleon scattering cross-section $\sigma_{DM,N}$ is given by:
\begin{equation}
 \sigma_{\mathrm{DM,N}}=\frac{\lambda_{345}^2}{4\pi M_h^4}\frac{m_N^4}{(m_N+M_{H})^2}f_N^2\, ,\label{directsigma}
\end{equation}
where we take $M_h=125$ GeV, $m_N=0.939$ GeV and $f_N=0.326$ as the universal Higgs-nucleon coupling.\footnote{There is no agreement on the value of the $f_N$ coupling and various estimations exist in the literature. Here we consider the  middle value of $0.14<f_N<0.66$ \cite{Andreas:2009hj}, and comment on the other possible values later in the text 
(see also discussion in~\cite{Goudelis:2013uca}).} Value of the $\lambda_{345}$ coupling is essential for the value of $\sigma_{DM,N}$ in the IDM and so we translate the limits for $\lambda_{345}$ obtained from $\rg$ measurements to $(M_H,\sigma_{DM,N})$ plane, used in direct detection experiments.

\begin{figure}[t]
\centering
\subfloat[$\rg>0.7$]{
\includegraphics[width=.45\textwidth]{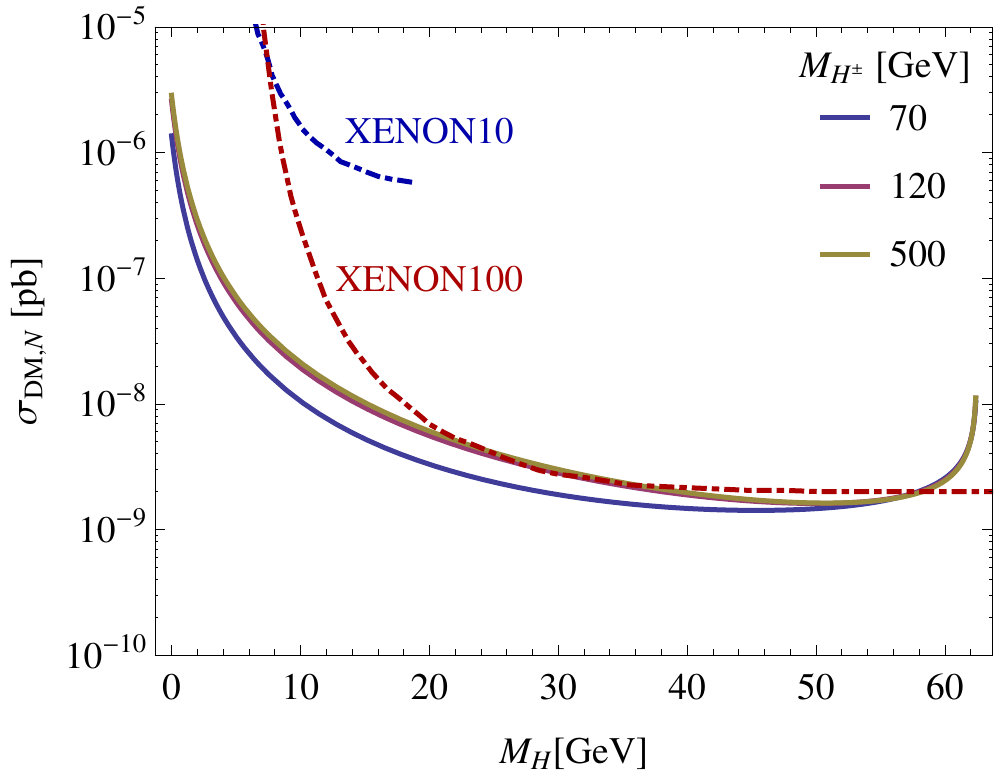}}
\subfloat[$\rg>0.8$]{
\includegraphics[width=.45\textwidth]{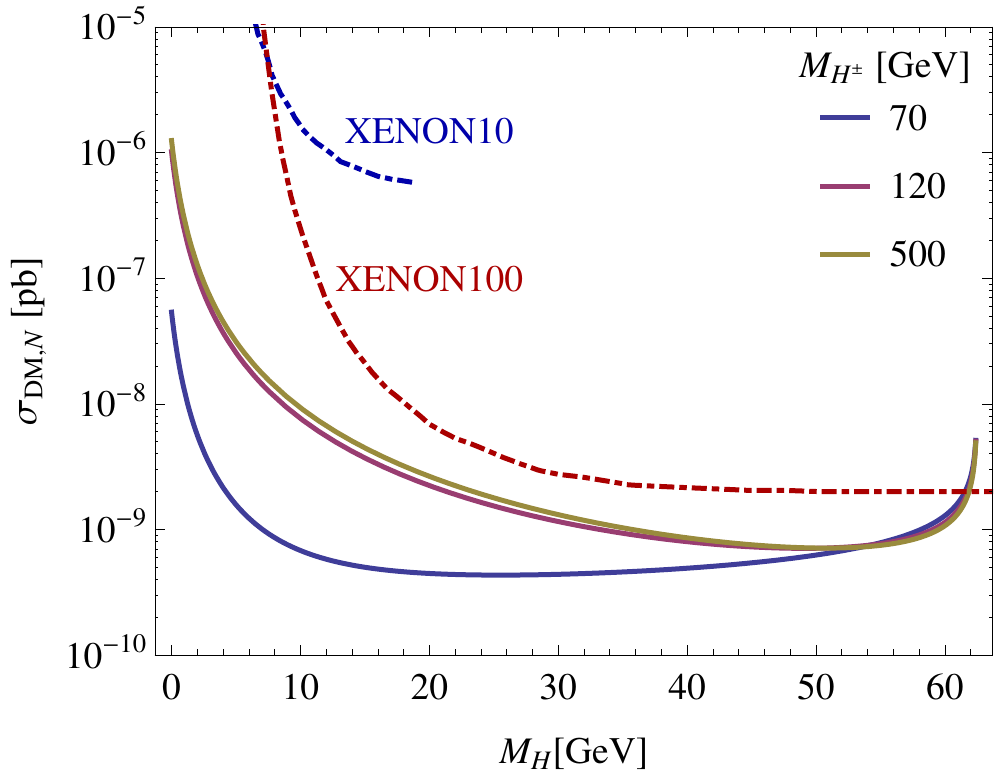}}
\caption{Upper limit on $\sigma_{\mathrm{DM},N}$ (\ref{directsigma}) with  $f_N = 0.326$ coming from the
requirement that (a) $\rg>0.7$, (b) $\rg>0.8$, expressed as a function of
$M_H$, for the case when the $h\to AA$ channel is closed.  Three values of
$M_{H^{\pm}}$ are considered: $\m=70\g$, $\m=120\g$, $\m=500\g$. For
comparison also the upper bounds set by XENON10 and XENON100 are shown.} \label{direct}
\end{figure}

Exclusion bounds for cases $\rg >0.7,\,0.8$ are shown in figure \ref{direct}, along with the XENON10/100 limits \cite{Aprile:2012nq}. If $H$ should constitute 100\% of DM in the Universe, then the limits set by $\rg$ measurements are  much stronger than those provided by XENON10/100 experiments for $M_H \lesssim 20$ GeV. Even  for  $\rg>0.7$ 
it provides stronger or comparable limits for $\sigma_{DM,N}$  for $M_H \lesssim 60$ GeV.

\subsection{Invisible decay channels closed}

If $M_H>M_h/2$, and consequently $M_A>M_h/2$, the invisible channels are closed and the only modification to $\rg$  comes from the charged scalar loop (\ref{Hloop}), so the most important parameters are $\m$ and $\lambda_{3}$ (or equivalently $m_{22}^2$). The contribution from the SM ($\mathcal{M}^{\mathrm{SM}}$) is real and negative and $\delta\mathcal{M}^{\mathrm{IDM}}$ is also real with sign correlated with the sign of $\lambda_3$. Enhancement in $\rg$ is possible when $\lambda_3<0$ \cite{Arhrib:2012, Swiezewska:2012eh, Krawczyk:2013gia, Krawczyk:2013wya}, with the maximal value of  $\rg$ approached for $\lambda_3=-1.47$, i.e. the smallest value of this parameter allowed by model constraints (\ref{constraints}).

The contribution to the amplitude from the charged scalar loop ($\delta\mathcal{M}^{\mathrm{IDM}}$) is a decreasing function of $M_{H^\pm}$ so in general the larger $\rg$ is, the smaller $\m$ should be. For example, $\rg>1.2$ gives $70\g<\m<154 \g$ \cite{Swiezewska:2012eh}.

\begin{figure}[t]
\centering
\subfloat[$\rg=1.01$]{
  \includegraphics[width=.46\textwidth]{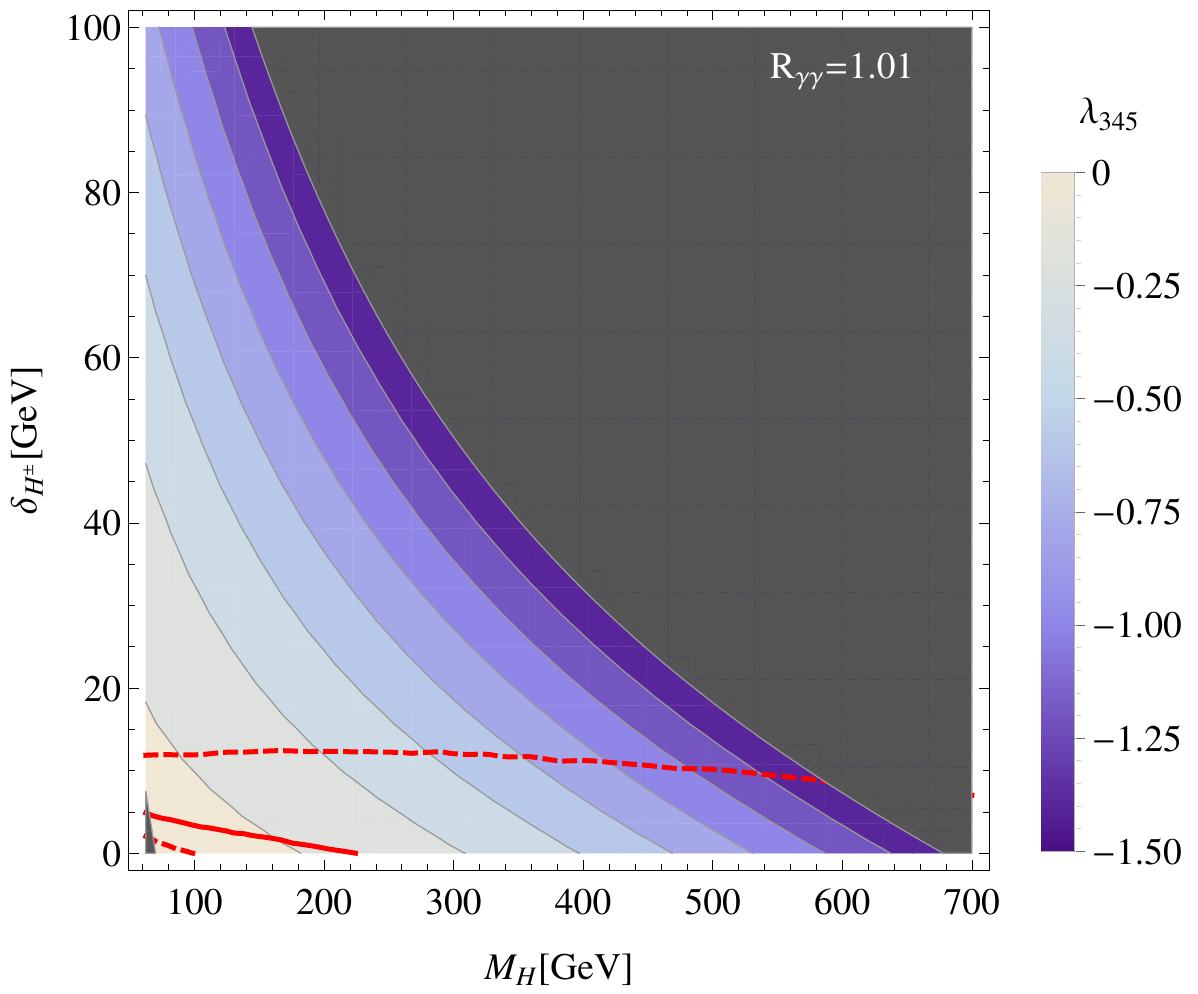}}
\subfloat[$\rg=1.02$]{  \includegraphics[width=.46\textwidth]{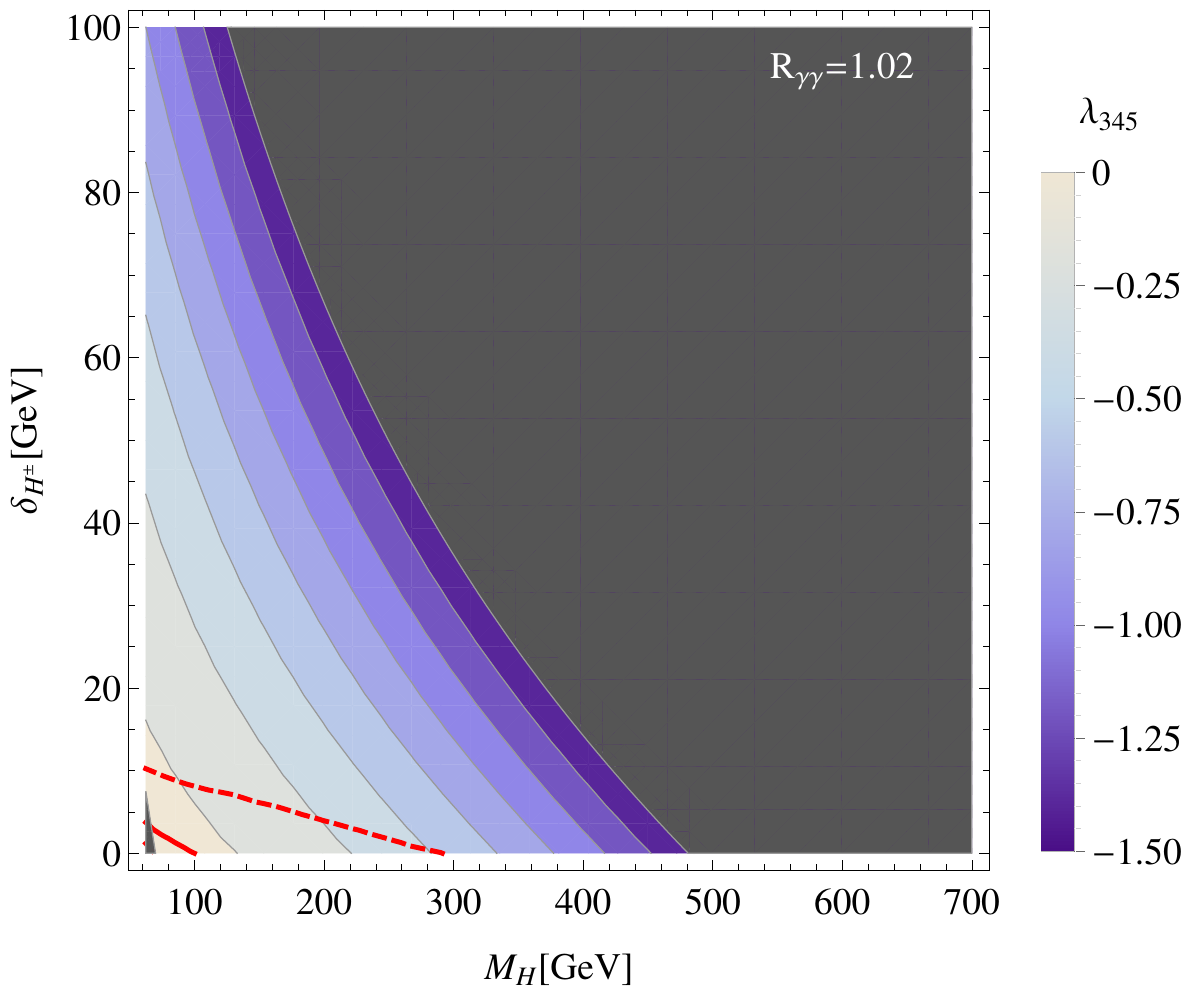}}
     \caption{Allowed regions in ($M_{H}$, $\delta_{H^{\pm}}$) plane for two values of $R_{\gamma\gamma}$: 1.01 (left panel), 1.02 (right panel). Dark grey region is excluded due to LEP bounds (left lower corner) and the vacuum stability/unitarity constraints (\ref{constraints}) (right upper corner). Red  lines show bounds from XENON100 (solid for $f_N=0.326$, dashed for $f_N=0.14$ and $f_N=0.66$) --- region above this line is excluded, if we assume that the dark scalar $H$ constitutes all dark matter relic density.}
     \label{rgg:fig:mhmap}
\end{figure}

Since for invisible channels closed $\rg$ depends only on $\m$ and $\lc$ (or $m_{22}^2$), fixing $\rg$ and $\m$ sets the value of $m_{22}^2$. For fixed $m_{22}^2$, $M_H$ depends only on $\lczp$,  eq.~(\ref{mass}). Thus, we can study the correlation between $\m$, $M_H$ and $\lczp$ for different values of $\rg$. 
Figure \ref{rgg:fig:mhmap} shows the ranges of $\lambda_{345}$ in the $(\m,\delta_{H^\pm})$ plane for two values of $\rg$ close to~1, $\rg = 1.01$ and $1.02$. One can see that even a small deviation from $\rg =1$ requires a relatively large $\lambda_{345}$, if the mass difference $\delta_{H^\pm}$ is of the order $(50-100)$ GeV. Small values of $|\lambda_{345}|$ are preferred if the mass difference is small.

\begin{figure}[t]
  \centering
  \includegraphics[width=.55\textwidth,]{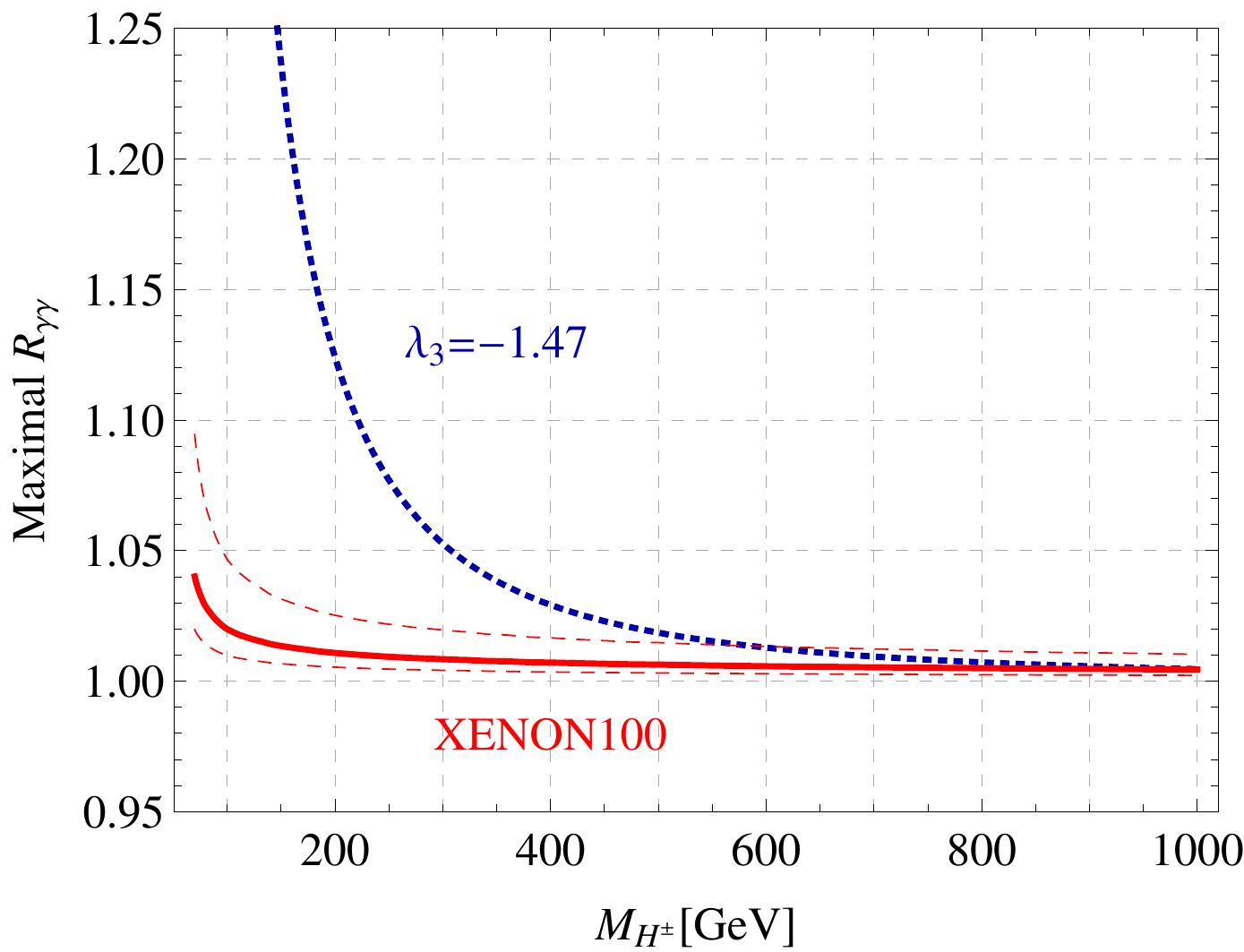}
     \caption{Blue (thick dashed) curve: the maximal value of $\rg$, allowed by the condition
$\lambda_{345}>-1.47$ (eq. (\ref{constraints})) as a function of $\m$.
Red lines (solid/thin dashed): the maximal value of $\rg$, allowed by the XENON100 constraints on
$\lambda_{345}$ (derived using the assumption that $H$ constitutes 100\% of
DM) as a function of $\m$, for $\m=M_H$. Solid red line corresponds to the
bounds obtained for $f_N = 0.326$ as in \cite{Djouadi:2011aa}, while upper dashed for $f_N = 0.14$
and lower dashed for $f_N = 0.66$.}
     \label{rgg:fig:rggmhp}
\end{figure}

Unitarity and positivity limits on $\lc$ and $\lambda_{345}$, eq.~(\ref{constraints}), constrain the allowed value of $\m$ (and thus also the mass of $H$) for a given value of $\rg$. For $\rg^\textrm{max}=1.01$ masses of $\m \gtrsim 700$~GeV are excluded, and if $\rg^\textrm{max} =1.02$ this bound is stronger, forbidding $\m \gtrsim 480$~GeV (figure \ref{rgg:fig:mhmap}).

The blue curve in figure~\ref{rgg:fig:rggmhp} shows the maximal value of $\rg$, which is obtained for maximally allowed negative value of $\lc=-1.47$, as a function of $\m$. In general, as previous studies have shown, the very heavy mass region is consistent with very small deviations from $\rg = 1$, but substantial enhancement of $\rg$ suggested by the central value measured by ATLAS (\ref{rg_atlas}) cannot be reconciled  with this region of masses.

$\rg <1$ is possible if the invisible channels are closed and $\lambda_3 >0$. Requiring that $\rg$ is bounded from above one can also limit the allowed parameter space. For example, if $\rg<0.8$ and invisible channels are closed, then $M_H < 200$ GeV~\cite{Swiezewska:2012eh}.

%

\paragraph{Comparison with XENON100 results}
If the dark scalars $H$ constitute 100\% of DM in the Universe, then the $\sigma_{DM,N}$ measurements done by the direct detection experiments bound the $\lambda_{345}$ parameter, which is also constrained by the $\rg$ value (figure \ref{rgg:fig:mhmap}). For given scalar masses one can test the compability between the two limits, and figure \ref{rgg:fig:mhmap} shows that $\rg >1$ and agreement with XENON100 need almost degenerated masses of $H$ and $H^\pm$. If $\delta_{H^\pm}$ is larger then $\rg$ requires larger $\lambda_{345}$, and that violates the XENON100 bounds (figure \ref{rgg:fig:mhmap}).

For a given value of $M_H$ one can find maximal negative value of $\lczp$ allowed by XENON100 experiment. Assuming that $M_H$ and $\m$ are degenerate allows to compute maximal  allowed value of $\rg$. The dependence of maximal $\rg$ on $\m=M_H$ for different values of $f_N$ is shown in figure~\ref{rgg:fig:rggmhp} (red curves).   For $M_H \approx M_{H^{\pm}}=70~\mathrm{GeV}$, the $R_{\gamma\gamma}$ is bounded by (1.09, 1.04, 1.02) for $f_N=$ (0.14, 0.326, 0.66) respectively. Thus it is  not possible to have $\rg>1.09$ in agreement with XENON100, unless the dark scalar $H$ constitutes only a part of the dark matter relic density. 

%


\section {Combining  $\rg$  and relic density constraints  on DM \label{consequences}}

In this section we compare the limits on the $\lambda_{345}$ parameter  obtained from $\rg$ in the previous section with those coming from the requirement that the DM relic density is in agreement with the WMAP measurements    (\ref{omega}). We use the micrOMEGAs package \cite{Belanger:2013oya} to calculate $\relic$ for chosen values of   DM masses. We demand that the obtained value lies in the 3$\sigma$ WMAP limit:
\begin{equation}
0.1018<\Omega_{DM} h^2<0.1234 \,. \label{WMAP}
\end{equation} 
If this condition is fulfilled, then $H$ constitutes 100\% of DM in the Universe. Values of $\Omega_{H}h^2>0.1234$ are excluded, while $\Omega_H h^2 <0.1018$ are still allowed if $H$ is a subdominant DM candidate.

\subsection{Low DM mass}

In the IDM the low DM mass region corresponds to the masses of $H$ below 10 GeV, while the other dark scalars are heavier, $M_A \approx M_{H^+} \approx 100$ GeV. In this region the main annihilation channel is $HH \to h \to \bar{b} b$ and to have the proper relic density, the $HHh$ coupling $(\lambda_{345})$ has to be large, above $\mathcal{O} (0.1)$. For example, for CDMS-II favoured mass $M = 8.6$ GeV \cite{Agnese:2013rvf} one gets relic density in agreement with bound (\ref{WMAP}) for $|\lambda_{345}| = (0.35-0.41)$, while $|\lambda_{345}| \lesssim 0.35$ are excluded. 

In the low mass region the invisible channel $h \to HH$ is open, meaning that $\rg>1$ is not possible, so we can conclude that $\rg>1$ (\ref{rg_atlas}) excludes the low DM mass region in the IDM. If $\rg<1$, as suggested by the CMS data (\ref{rg_cms}), the low DM mass could be in principle allowed. However, our results, described in the previous section, show, that it is not possible, as the  coupling allowed by $\rg$, i.e. $|\lambda_{345}| \sim 0.02$, is of an order of magnitude smaller than needed for $\relic$. So we can conclude that the low DM mass region cannot be accommodated in the IDM with recent LHC results, irrespective of whether $H$ is the only, or just a subdominant, DM candidate.

\subsection{Medium DM mass}

\begin{figure}[t]
\centering
\includegraphics[width=.55\textwidth]{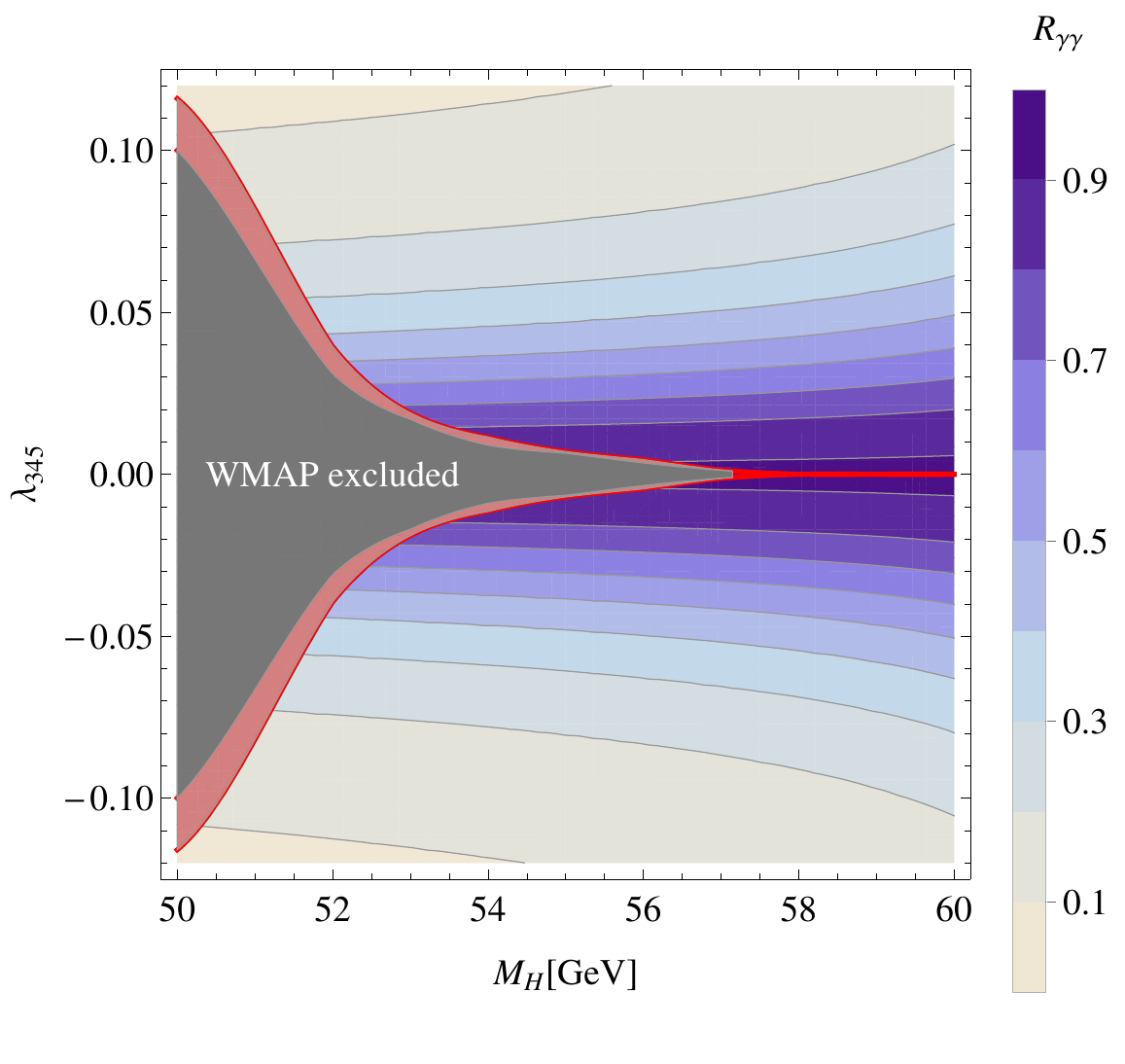}
\caption{Comparison of the values of $\rg$ and region allowed by the relic density measurements for the middle DM mass region with $HH$ invisible channel open and $M_A = M_{H^\pm} = 120$ GeV. Red bound: region in agreement with WMAP (\ref{WMAP}). Grey area: excluded by WMAP. $\rg >0.7$ limits the allowed values of masses to $M_H > 53$ GeV.} \label{midOmega}
\end{figure}

\paragraph{Invisible decay channels open}
Let us first consider the case with  $AA$ invisible channel closed, where we chose $M_A = M_{H^\pm} = 120$ GeV. In this case the main annihilation channels are $HH \to h \to \bar{f}f$, when the $HHh$ coupling is large enough and $HH \to W^+ W^-$, when the $HHh$ coupling is suppressed, typically leading to $\relic$ above the WMAP limit. Lower values of $M_H$ require rather large $\lambda_{345}$ --- in this sense this region resembles the low DM mass region. As $M_H$ grows towards $M_H = M_h/2$, the value of $\lambda_{345}$ required to obtain the proper relic density gets smaller, leading eventually to the $\relic$ below WMAP limit, apart from extremely tunned and small values of $\lambda_{345}$.

These results are presented in figure \ref{midOmega}, where the WMAP-allowed range of $\relic$ is denoted by the red bound. Grey excluded region between the WMAP bounds corresponds to $\relic$ too large, leading to the overclosing of the Universe. If we consider $H$ as a subdominant DM candidate with $\Omega_H h^2 < \relic$ then also the regions below and above  red bounds in figure \ref{midOmega} are allowed. This usually corresponds to the larger values of $\lambda_{345}$. 
It can be clearly seen that for a large portion of the parameter space limits for $\lambda_{345}$ from $\rg$, even for the least stringent case $\rg >0.7$, cannot be reconciled with the WMAP-allowed region.

\begin{figure}[t]
\centering
\includegraphics[width=.55\textwidth]{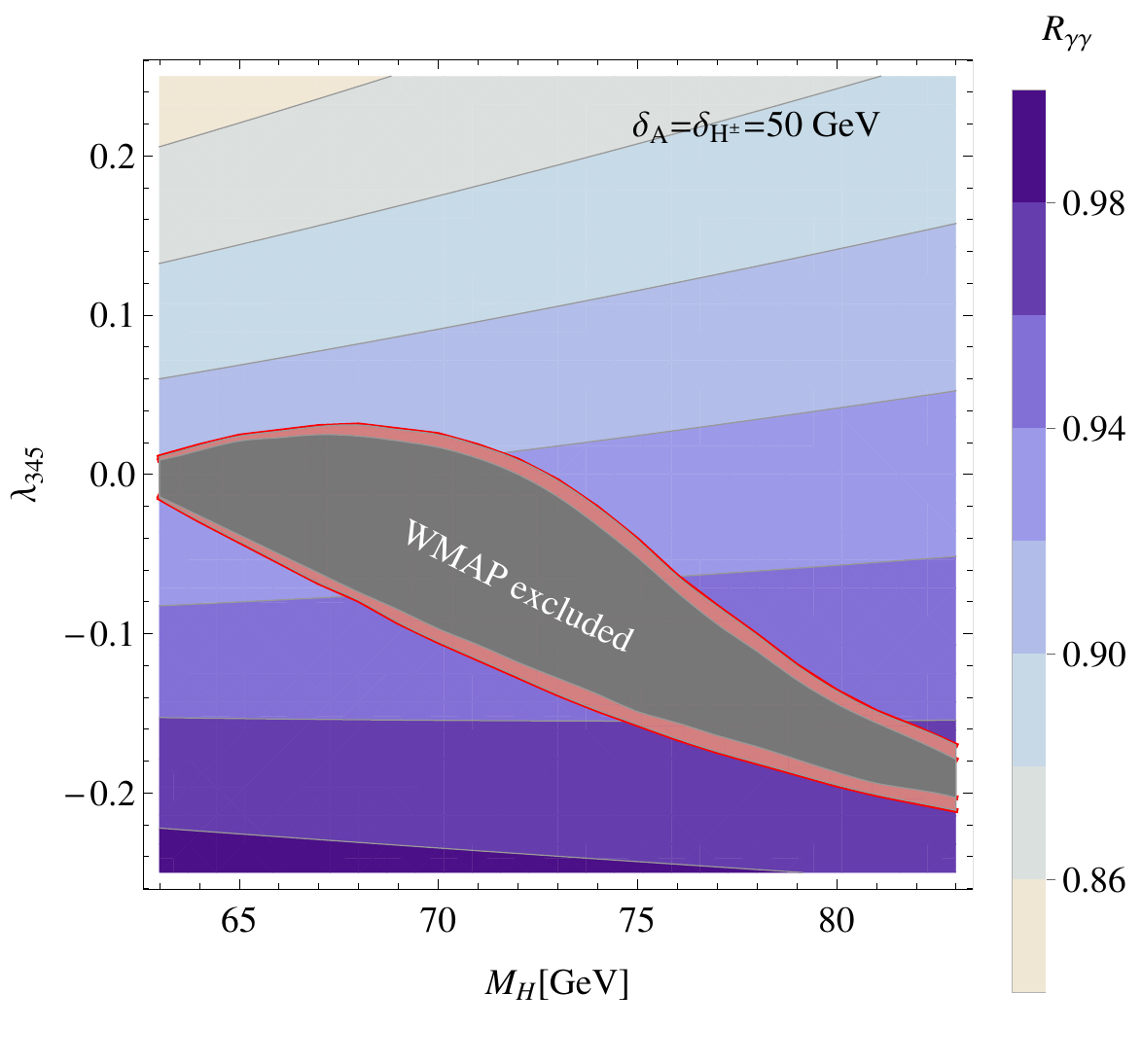}
\caption{Comparison of the values of $\rg$ and region allowed by the relic density measurements for the middle DM mass region with $HH$ invisible channel closed and $\delta_A = \delta_{H^\pm} = 50$ GeV. Red bound: region in agreement with WMAP (\ref{WMAP}). Grey area: excluded by WMAP. $\rg >1$ is not possible, unless $H$ is a subdominant DM candidate.} \label{midOmega2}
\end{figure}

\paragraph{Invisible decay channels closed} In this analysis we choose $\delta_{H^\pm} = \delta_A = 50$ GeV in agreement with the set of constraints (\ref{LEP}) and $M_H$ varying between $M_h/2$ and $83$ GeV. The main annihilation channels are as in the previous case, with the gauge channels getting more important as the mass of the DM particle grows. This, and the 
presence of the three body final states with virtual $W^\pm$, are the main reason why the WMAP-allowed region (the red bound) presented in figure \ref{midOmega2} is not symmetric around zero, eventually leaving no positive values of $\lambda_{345}$ allowed. The absolute values of $\lambda_{345}$ that lead to the proper relic density are in general larger than in the case of $M_H<M_h/2$.

Figure \ref{midOmega2} presents the values of $\rg$ for chosen masses and couplings compared to the WMAP-allowed/excluded region. It can be seen that this region is consistent with $\rg<1$. It is in agreement with results obtained before (figure \ref{rgg:fig:mhmap}), as mass difference $\delta_{H^\pm} = 50$ GeV and $\rg>1$ requires $\lambda_{345} \lesssim -0.3$, a value smaller than the one obtained from the relic density limits.

We can conclude, that $\rg>1$ and relic density constraints (\ref{WMAP}) cannot be fulfilled for the middle DM mass region. If the IDM is the source of all DM in the Universe and $M_H \approx(63-83)$ GeV then the maximal value of $\rg$ is around $0.98$. A subdominant DM candidate, which corresponds to larger $\lambda_{345}$, is consistent with $\rg>1$. 

\paragraph{Comparison with the indirect detection limits} 

The current best limits on DM annihilation into $b \bar{b}$, which is the main annihilation channel for
low and medium DM masses in the IDM, come from the measurements of  secondary photons from Milky Way dwarf galaxies and the Galactic Centre region by the Fermi-LAT satellite \cite{Ackermann:2011wa, Ackermann:2012qk}.\footnote{AMS-02 results provide weaker constraints on Dark Matter annihilation into $b \bar{b}$ \cite{Kopp:2013eka}.} They exclude the generic WIMP candidates that annihilate mainly into $b \bar{b}$ and reproduce the observed $\Omega_{DM} h^2$ for $M_{DM} \lesssim 25$ GeV \cite{Ackermann:2011wa}. Independent analyses give slightly stronger limits, excluding generic WIMPs with $M_{DM}$ less than $40$ GeV  \cite{Abazajian:2011tk,GeringerSameth:2011iw}. Observations of  $\gamma$ line signals give no further constraints for standard WIMP models \cite{Bringmann:2012ez}. Indirect detections exclusions are in general weaker than those provided by XENON10/XENON100 experiments and can be comparable or stronger only in the low mass regime, where the controversies from the direct detection are the strongest. The combined $\relic$ and $\rg$ analysis, performed here, excludes masses of DM in the IDM below $53$ GeV if $\rg >0.7$ and thus gives stronger limits on the allowed values of masses in the IDM than those currently obtained from the indirect detection experiments.

\section{Summary}

The IDM is a simple extension of the Standard Model that can provide a scalar DM candidate. This candidate is consistent with the WMAP results on the DM relic density and in three regions of masses it can explain 100~\% of the DM in the Universe. In a large part of the parameter space it can also be considered as a subdominant DM candidate.
Measurements of the diphoton ratio, $\rg$, recently done by the ATLAS and the CMS experiments at the LHC, set strong limits on masses of the DM and other dark scalars, as well as the self-couplings, especially $\lambda_{345}$. In this paper we discuss the obtained constraints for various possible values of $\rg$, that are in agreement with the recent LHC measurements,  and combine them with WMAP constraints. 

  The main results of the present paper are as follows:
\begin{itemize}
\item
If invisible Higgs decays channels are open ($M_{H}<M_h/2$) then $\rg$ measurements can constrain the maximal value of $|\lambda_{345}|$. This sets strong limits especially on the low DM mass region in the IDM. Values of $|\lambda_{345}|$ that lead to the proper relic density in the 3$\sigma$ WMAP range $0.1018<\Omega_{DM} h^2<0.1234$, 
are an order of magnitude larger than those allowed by assuming that $\rg>0.7$. We conclude that we can exclude the low DM mass region in the IDM, i.e. $M_H \lesssim 10$ GeV.
\item
$\rg$ also provides  strong limits for  larger values of $M_H$. First, demanding that $\rg >0.9$ leaves only a~small part of the parameter space allowed, excluding the region $M_A - M_H \gtrsim 2$ GeV if both invisible decay channels are open or $M_H \lesssim 43$ GeV if the $AA$ channel is closed. Second, comparing $\rg$ limits with the WMAP allowed region, we found that masses $M_H \lesssim 53$ GeV, which require larger values of $\lambda_{345}$ to be in agreement with WMAP, cannot be reconciled with $\rg>0.7$.
\item
$\rg$ sets limits on the DM-nucleon scattering cross-section in the low and medium DM mass region, 
which are stronger or comparable with the results obtained both by the XENON100 and Fermi-LAT experiments.
\item
If the invisible decay channels are closed, then $\rg >1$ is possible. This however leads to the constraints on masses and couplings. In general, $\rg>1$ favours the degenerated $H$ and $H^\pm$. When the mass difference is large, $\delta_{H^\pm} \approx (50-100)$ GeV, then the required values of $|\lambda_{345}|$ that provide $\rg>1$ are bigger than those allowed by WMAP measurements. We conclude it is not possible to have all DM in the Universe explained by the IDM (in the low and medium DM mass
regime) and $\rg>1$. If $\rg>1$ then $H$ may be a subdominant DM candidate. If $\rg <1$ then $M_H\approx (63-80)$ GeV can explain 100\% of DM in the Universe.
\end{itemize}


\paragraph{Acknowledgments} We thank Sabine Kraml and Sara Rydbeck for comments and suggestions. This work was supported in part by the grant  NCN OPUS 2012/05/B/ST2/03306 (2012-2016).


\end{document}